\documentclass[conference]{IEEEtran}
\usepackage{amsmath}
\usepackage[pdftex]{graphicx}
\usepackage{graphicx}
\usepackage{pstricks}
\usepackage{amsfonts}
\usepackage{tabularx}
\newtheorem{Lemma}{Lemma}
\newtheorem{Remark}{Remark}

\usepackage{amssymb}
\usepackage{multirow}
\usepackage{epsfig}
\usepackage{epstopdf}

\usepackage[noadjust]{cite} 

\begin{document}
\title{Untrusted NOMA with Imperfect SIC: Outage Performance Analysis and Optimization}
\author{\IEEEauthorblockN{Sapna Thapar$^{1}$, Deepak Mishra$^{2}$, and Ravikant Saini$^{1}$}
\IEEEauthorblockA{$^{1}$Department of Electrical Engineering, Indian Institute of Technology Jammu, India\\
$^{2}$School of Electrical Engineering and Telecommunications, University of New South Wales, Australia\\
Emails: thaparsapna25@gmail.com, d.mishra@unsw.edu.au, ravikant.saini@iitjammu.ac.in
}}
\maketitle

\begin{abstract}
Non-orthogonal multiple access (NOMA) has come
to the fore as a spectral-efficient technique for fifth-generation
and beyond communication networks. We consider the downlink
of a NOMA system with untrusted users. In order to consider
a more realistic scenario, imperfect successive interference
cancellation is assumed at the receivers during the decoding
process. Since pair outage probability (POP) ensures a minimum
rate guarantee to each user, it behaves as a measure of the quality
of service for the pair of users. With the objective of designing
a reliable communication protocol, we derive the closed-form
expression of POP. Further, we find the optimal power allocation that minimizes
the POP. Lastly, numerical results have been presented which
validate the exactness of the analysis, and reveal the effect of
various key parameters on achieved pair outage performance.
In addition, we benchmark optimal power allocation against
equal and fixed power allocations with respect to POP. The
results indicate that optimal power allocation results in improved
communication reliability.
\end{abstract}

\section{Introduction}
Non-orthogonal multiple access (NOMA) has been recognized as a disruptive technology, supporting massive connections through high spectral efficiency, for fifth-generation and beyond wireless networks. With NOMA, a number of users are accommodated simultaneously within the same resource block by using superposition coding at the transmitter and successive interference cancellation (SIC) at the receiver \cite{nomasurvey}. However, each user in the system has a threshold data rate demand, i.e., each user desires the base station to transmit messages with a minimum data rate guarantee under the quality of service (QoS) agreement. Therefore, ensuring reliability with a minimum rate guarantee to each user in an untrusted NOMA system is an interesting research direction. 

%From the secrecy perspective, NOMA is confronted with security issues due to wireless transmission and SIC at receiver, which gets critically significant in the presence of untrusted users. Therefore, ensuring reliability along with security in an untrusted NOMA system an interesting research direction. 

%Secure communications in an untrusted NOMA system means that the users   sharing the same resource block in NOMA, dont trust each other. Therefore, the data of each user should be safeguarded from all other users. Further, an untrusted system is a hostile but practical situation  \cite{saini2016ofdma}. 

\subsection{Related Works}
Reliable communication signifies that the messages broadcasted from the transmitter are highly likely to reach their destination. Extensive research activities have analyzed the reliability in terms of outage probability for the considered different system models. In \cite{6868214}, the performance analysis of a downlink NOMA system with randomly deployed mobile users for a Rayleigh fading channel link was investigated in terms of outage probability. In \cite{8272349}, outage performance analysis for a two-user NOMA system with Nakagami-m fading channel link was studied. In \cite{8063934}, closed-form outage probability for each user in a NOMA system with statistical channel state information at the transmitter and Nakagami-m fading links was investigated. The exact expressions for the outage probability for a two-user
NOMA system considering a dynamic power allocation scheme was investigated in \cite{7542118}. In the direction of an untrusted NOMA system, reliability was discussed in \cite{basepaper}, where the authors' derived pair outage probability (POP) to analyze the reliability of a two-user NOMA system. Then, in \cite{9188014}, POP was analyzed for the optimal decoding order that can ensure secrecy to each user of the system.

\subsection{Motivation and Key Contributions}
The outage analyses studied in the above-mentioned works \cite{6868214}-\cite{basepaper} are restricted to the conventional decoding order of NOMA. Unlike the conventional decoding order, each user can perform SIC and decode data of itself and other users in any sequence if the users are untrusted \cite{9188014}, \cite{9324786}. As a result, a number of decoding orders are possible in an untrusted NOMA system \cite{9324786} and each of them is important for different applications. Thus, it is important to study the outage performance in each decoding order. Besides, a common assumption considered in \cite{6868214}-\cite{basepaper} is that perfect SIC is performed by the receivers. With perfect SIC, the interference from the decoded users is cancelled entirely while decoding later users which results in better spectral efficiency. Nevertheless, implementing perfect SIC might not be practical, due to decoding errors and complexity scaling issues \cite{7881111}, \cite{8755843}. In consequence, imperfect SIC, where the residual interference (RI) from imperfectly decoded users remains while decoding later users should be considered \cite{8755843}, \cite{eusipco}. In the NOMA literature, some researchers have considered a fixed value of RI \cite{constant_sic1}, \cite{constant_sic2} while many have taken RI as a linear function of the interfering power \cite{7881111}, \cite{8755843}, \cite{letter}. However, despite its significance, analyzing the outage performance of an untrusted NOMA system with imperfect SIC has not received much attention yet. 

Assuming imperfect SIC at receivers, POP of the system was analyzed with RI in \cite{9188014}. Here, considering the application of secure communication in NOMA, the study was carried out for that optimal decoding order which ensures positive secrecy to each user. Note that in this work, a fixed RI value was considered, which is a strong and unrealistic assumption. In contrast to this, a more generalised linear model of imperfect SIC can effectively represent the relationship between RI and the received signal power. \emph{Therefore, to appreciate the significant impact of imperfect SIC with linear RI model on reliable communication for the users, we investigate the pair outage for a two-user NOMA system, which to the best of our knowledge, has not been investigated yet in the literature.}

The key contributions of this work are summarized below: 
\begin{itemize}
\item To observe the realistic impact of imperfect SIC with linear model, the analytical expressions of POP for a two-user untrusted NOMA system is derived.
\item To analyze optimal performance, optimal power allocation minimizing POP is obtained by using the concept of generalized-convexity of POP.
\item Numerical results are provided to validate the analytical results, and highlight deeper insights on the impact of different key parameters on the system performance.
\end{itemize} 

\section{Untrusted NOMA with Imperfect SIC}
In this section, we first present the system model, then provide the possible decoding orders for an untrusted NOMA system. Finally, the signal-to-interference-plus-noise-ratio (SINR) expressions are derived for users under imperfect SIC scenario.  

\subsection{System Model}
We consider downlink transmission of a NOMA system consisting of one base station (BS) and two untrusted users. It is not preferable in practice to ask all users in the system to participate in NOMA jointly due to an increase in co-channel interference and implementation complexity with more users. Therefore, dividing users into groups and implementing NOMA in each group is a better choice. For analytical tractability, we focus on a two-user NOMA system as taken in many works \cite{ding2016impact}, \cite{8309422}. The $n$-th user is denoted by $U_{n}$, where $n\in \mathcal{N}=\{1, 2\}$. All the nodes in the network are assumed to be equipped with one antenna. The channel gain from BS to $U_{n}$ is assumed to be Rayleigh distributed. We denote the channel gain coefficient between BS and $U_{n}$ by  $h_{n}$. The channel power gain $|h_{n}|^{2}$ follows an exponential distribution with mean parameter $\lambda_{n}=L_{p}d_{n}^{-e}$, where $d_{n}$ indicates the distance between BS and $U_{n}$, $L_{p}$ denotes the path loss constant, and $e$ is the path loss exponent. Without loss of generality, channel power gains are assumed tio be ordered as $|h_{1}|^{2}>|h_{2}|^{2}$. Therefore, near and far user, i.e., $U_{1}$ and $U_{2}$ could be regarded as strong and weak user, respectively. 

In accordance to the power-domain NOMA principle, the BS transmits the superimposed signal $\sqrt{\alpha P_{t}} x_{1}$+ $\sqrt{(1-\alpha)P_{t}} x_{2}$ to both users. Here, $x_{1}$ and $x_{2}$ are the information signals associated with $U_1$ and $U_2$, respectively. $P_{t}$ is the total BS transmission power. $\alpha$ denotes the fraction of $P_{t}$ allocated to $U_{1}$, and $(1-\alpha)$ is the remaining fraction of $P_{t}$ allocated to $U_{2}$. At the receiver side, each user performs SIC where inter-user interference is cancelled out to extract the desired information signal \cite{7343355}. Without loss of generality, received additive white Gaussian noise is assumed to have mean equal to zero and variance equal to $\sigma^{2}$ at both the users. In order to focus on a realistic system model, we consider imperfect SIC at receivers where RI from imperfectly decoded signals remains while decoding later users. Considering linear RI model, the corresponding RI factor is denoted by $\beta$, where $0\leq\beta\leq1$. Note that $\beta=0$ corresponds to perfect SIC and $\beta=0$ indicates fully imperfect SIC, i.e., maximum interference \cite{7881111}, \cite{8755843}.

\subsection{Possible Decoding Orders for Untrusted NOMA}\label{decodingorders}
Each receiver performs SIC in a certain sequence. A collection of such sequences for each of the users is termed as the ``\textit{decoding order}'' of the system. In accordance with an untrusted NOMA system, each user can decode any user's signal at any stage \cite{9188014}, \cite{9324786}, \cite{7343355}.  Thus, in a two user system, with each user having two choices that of decoding self or the other user first, four decoding orders are possible \cite{9188014}. We can denote the decoding order as a $2\times 2$ matrix $\mathbf{D}_{o}$, where $o \in \{1,2,3,4\}$ indicates the index of $o$-th decoding order. The $m$-th column of matrix $\mathbf{D}_{o}$ is denoted by a column vector $\mathbf{d}_{m}$ of size $2\times1$, which imply the SIC sequence followed by $U_{m}$, where $m \in \mathcal{N}$.  Here, $[\mathbf{d}_{m}]_{k}=n$ signifies that $U_{m}$ decodes data of $U_{n}$ at $k$-th stage, where $n, k \in \mathcal{N}$ and $[\mathbf{d}_{m}]_{1} \neq [\mathbf{d}_{m}]_{2}$. 
Thus, we can write the four possible decoding orders as $\mathbf{D}_{1}=[2,1;2,1]$, $\mathbf{D}_{2}=[2,1;1,2]$, $\mathbf{D}_{3}=[1,2;2,1]$, and $\mathbf{D}_{4}=[1,2;1,2]$. 
\begin{Remark}
From the perspective of ensuring secure communication to both the users, it has been proved in \cite[Theorem 2]{9188014} that the optimal decoding order with respect to providing maximum secrecy rate for both users is $\mathbf{D}_{2}$. Therefore, for mathematical tractability, we further derive the analytical expression of POP with the linear imperfect SIC model for decoding order $\mathbf{D}_{2}$. However, in a similar manner, we will be analyzing the POP expressions also for other decoding orders in the extended version of this work.
\end{Remark}

\subsection{Achievable SINRs with Imperfect SIC}
According to the decoding order $\mathbf{D}_{2}$, $U_{1}$ and $U_{2}$ decode signal of other user at the first stage, and then decode their own signal at the second stage after performing SIC \cite{9188014}. Thus, considering imperfect SIC at receivers with linear model  \cite{9324786}, \cite{8755843}, the achievable SINR $\Gamma_{nm}$ at $U_{m}$, when $U_{n}$ is decoded by $U_{m}$,  where $m, n \in \mathcal{N}$, is given as 
\begin{align}
\Gamma_{21} &= \frac{(1-\alpha)|h_{1}|^{2}}{\alpha|h_{1}|^{2}+\frac{1}{\rho_{t}}}, \\
\Gamma_{12} &= \frac{\alpha|h_{2}|^{2}}{(1-\alpha)|h_{2}|^{2} + \frac{1}{\rho_{t}}},   \\
\Gamma_{11} &= \frac{\alpha|h_{1}|^{2}}{(1-\alpha)\beta|h_{1}|^{2}+\frac{1}{\rho_{t}}}, \\
\Gamma_{22} &= \frac{(1-\alpha)|h_{2}|^{2}}{\alpha\beta|h_{2}|^{2}+\frac{1}{\rho_{t}}},
\end{align}
where $\rho_{t}\stackrel{\Delta}{=}\frac{P_{t}}{\sigma^{2}}$ is the BS transmit signal-to-noise ratio (SNR). The corresponding achievable data rate at $U_{n}$ is given by the Shannon's capacity formula as
\begin{align}
R_{nm} = \log_{2}(1+\Gamma_{nm}).
\end{align} 

\section{Pair Outage Performance Analaysis}\label{sectionIV}
In this section, we derive the analytical expressions of POP that ensures users' QoS demands for reliable communication over both the links.  POP is defined as the probability at which the achievable data rate at each user falls below than the required threshold data rate. Thus, POP ensures minimum data rate guarantee to each user in the system. Let us denote POP for the system by $P_o$. Assuming threshold data rate for $U_{n}$ as $R_{n}^{\mathrm{th}}$, POP can be mathematically given as
\begin{align}\label{POP}
 P_{o} &= 1 - \mathbb{P}\{\Gamma_{11}>\pi_{1}, \Gamma_{21}>\pi_{2}, \Gamma_{12}>\pi_{1}, \Gamma_{22}>\pi_{2}\},\nonumber \\
& \stackrel{(\mathrm{g})}{=} 1 - \mathbb{P}\{\Gamma_{11}>\pi_{1}, \Gamma_{21}>\pi_{2}\} \mathbb{P}\{ \Gamma_{12}>\pi_{1}, \Gamma_{22}>\pi_{2}\},\nonumber \\ 
& = 1 -\mathbb{P}\{|h_{1}|^{2}> \max(\zeta_{1},\zeta_{2})\} \mathbb{P}\{|h_{2}|^{2}> \max(\zeta_{3},\zeta_{4})\},\nonumber \\
& = 1 - (1 - F_{|h_{1}|^{2}}(\max(\zeta_{1},\zeta_{2}))) (1 - F_{|h_{2}|^{2}}(\max(\zeta_{3},\zeta_{4}))),\nonumber \\
& = 1 - \bar{F}_{|h_{1}|^{2}}(\max(\zeta_{1},\zeta_{2})) \bar{F}_{|h_{2}|^{2}}(\max(\zeta_{3},\zeta_{4})),
\end{align}
where $\mathbb{P}\{.\}$ denotes the probability measure, $\pi_{n} \stackrel{\Delta}{=} 2^{R_{n}^{\mathrm{th}}} - 1$, $(\mathrm{g})$ follows from the property of independent events \cite{papoulis2002probability}, $\zeta_{1}\stackrel{\Delta}{=} \frac{\pi_{1}}{(\alpha-\beta(1-\alpha)\pi_{1})\rho_{t}}, \zeta_{2}\stackrel{\Delta}{=}\frac{\pi_{2}}{(1-\alpha-\alpha\pi_{2})\rho_{t}}, \zeta_{3} \stackrel{\Delta}{=}\frac{\pi_{1}}{(\alpha-(1-\alpha)\pi_{1})\rho_{t}}$, and $\zeta_{4} \stackrel{\Delta}{=}\frac{\pi_{2}}{(1-\alpha-\beta\alpha\pi_2)\rho_{t}}$. $F_{|h_{1}|^{2}}(x)$ and $\bar{F}_{|h_{1}|^{2}}(x)$, respectively, denote the cumulative distribution function (CDF) and complementary cumulative distribution function (CCDF), of channel power gain $|h_{1}|^{2}$. In a similar manner, $F_{|h_{2}|^{2}}(x)$ and $\bar{F}_{|h_{2}|^{2}}(x)$ are CDF and CCDF, respectively, of channel power gain $|h_{2}|^{2}$. 

We observe that in \eqref{pair_outage}, $\bar{F}_{|h_{1}|^{2}}(\max(\zeta_{1},\zeta_{2}))$ can be rewritten for two cases $\zeta_{1} >\zeta_{2}$ and $\zeta_{1}<\zeta_{2}$. On solving the first case $\zeta_{1} >\zeta_{2}$, we find a constraint on $\alpha$ as $\alpha < \frac{\pi_{1}(1+\pi_{2}\beta)}{\pi_{2}(1+\pi_{1}\beta)+\pi_{1}(1+\pi_2)}$. Similarly, in the second case $\zeta_{1} < \zeta_{2}$, the constraint on $\alpha$ will be $\alpha > \frac{\pi_{1}(1+\pi_{2}\beta)}{\pi_{2}(1+\pi_{1}\beta)+\pi_{1}(1+\pi_2)}$. Besides, from the definition of CDF of exponential distribution, $\zeta_{1}>0$ and $\zeta_{2}>0$, respectively, gives $\alpha>\frac{\beta\pi_1}{1+\beta\pi_1}$ and $\alpha<\frac{1}{1+\pi_{2}}$. Thus, denoting $\alpha_1\stackrel{\Delta}{=}\frac{\beta\pi_1}{1+\beta\pi_1}$, $\alpha_{2}\stackrel{\Delta}{=}\frac{\pi_{1}(1+\pi_{2}\beta)}{\pi_{2}(1+\pi_{1}\beta)+\pi_{1}(1+\pi_2)}$, and $\alpha_{3} \stackrel{\Delta}{=}\frac{1}{1+\pi_{2}}$, $\bar{F}_{|h_{1}|^{2}}(\max(\zeta_{1},\zeta_{2}))$ can be expressed as
\begin{align}\label{pair_outage1}
\bar{F}_{|h_{1}|^{2}}(\max(\zeta_{1},\zeta_{2}))=     
\begin{cases}
 \exp\{-\frac{\zeta_{1}}{\lambda_{1}}\},  & \alpha_1<\alpha<\alpha_{2} \\
 \exp\{-\frac{\zeta_{2}}{\lambda_{1}}\},  & \alpha_{2}<\alpha<\alpha_{3} \\
  0, & \text{otherwise}.    
\end{cases}  
\end{align}

Similarly, $\bar{F}_{|h_{2}|^{2}}(\max(\zeta_{3},\zeta_{4}))$ can be expressed as follows
\begin{align}\label{pair_outage2}
\bar{F}_{|h_{2}|^{2}}(\max(\zeta_{3},\zeta_{4})) &=
\begin{cases}
  \exp\{-\frac{\zeta_{3}}{\lambda_{2}}\},  & \alpha_{4}<\alpha\!<\alpha_{5}\\
   \exp\{-\frac{\zeta_{4}}{\lambda_{2}}\}, & \alpha_{5}<\alpha\!<\alpha_{6}\\
           0, & \text{otherwise}
      \end{cases}
\end{align}
where $\alpha_{4}=\frac{\pi_{1}}{1+\pi_{1}}$, $\alpha_{5}=\frac{\pi_{1}(1+\pi_{2})}{\pi_{2}(1+\pi_{1})+\pi_{1}(1+\beta\pi_{2})}$, and $\alpha_{6}=\frac{1}{1+\beta\pi_{2}}$. Here we observe that $\alpha_{4}>\alpha_{1}$ and $\alpha_{6}>\alpha_{3}$ since $\beta<1$. Thus, considering \eqref{POP},  \eqref{pair_outage1}, \eqref{pair_outage2}, and $\alpha_{4}>\alpha_{1}$, $\alpha_{6}>\alpha_{3}$, the piecewise definition of $P_{o}$ as a function of $\alpha$ can be obtained as given at the top of next page in \eqref{pair_outage}.
\begin{figure*}
\begin{equation}\label{pair_outage}
P_{o} = %
\begin{cases}
1-\bar{F}_{|h_{1}|^{2}}(\zeta_{1})  \bar{F}_{|h_{2}|^{2}}(\zeta_{3}), \quad \quad \begin{cases} 
[\alpha_{4} < \alpha < \alpha_{5}]  \wedge [ \alpha_{5} < \alpha_{2}] \\
[\alpha_{4} < \alpha < \alpha_{2}] \wedge [ \alpha_{5} > \alpha_{2}] \\
\end{cases}
&\quad \quad \text{Case 1}
\\
1-\bar{F}_{|h_{1}|^{2}}(\zeta_{1})  \bar{F}_{|h_{2}|^{2}}(\zeta_{4}), \quad \quad \begin{cases} 
[\alpha_{5} < \alpha < \alpha_{6}]  \wedge [ \alpha_{5} > \alpha_{1}] \wedge [\alpha_{6} < \alpha_{2}]\\
[\alpha_{1} < \alpha < \alpha_{2}] \wedge [ \alpha_{5} < \alpha_{1}] \wedge [\alpha_{6} > \alpha_{2}]\\
[\alpha_{1} < \alpha < \alpha_{6}] \wedge [ \alpha_{5} < \alpha_{1}] \wedge [\alpha_{6} < \alpha_{2}]\\
[\alpha_{5} < \alpha < \alpha_{2}] \wedge [ \alpha_{5} > \alpha_{1}] \wedge [\alpha_{6} > \alpha_{2}]\\
\end{cases}
&\quad \quad \text{Case 2}
\\
1-\bar{F}_{|h_{1}|^{2}}(\zeta_{2})  \bar{F}_{|h_{2}|^{2}}(\zeta_{3}), \quad \quad
\begin{cases} 
[\alpha_{4} < \alpha < \alpha_{5}]  \wedge [ \alpha_{4} > \alpha_{2}] \wedge [\alpha_{5} < \alpha_{3}]\\
[\alpha_{2} < \alpha < \alpha_{3}] \wedge [ \alpha_{4} < \alpha_{2}] \wedge [\alpha_{5} > \alpha_{3}]\\
[\alpha_{2} < \alpha < \alpha_{5}] \wedge [ \alpha_{4} < \alpha_{2}] \wedge [\alpha_{5} < \alpha_{3}]\\
[\alpha_{4} < \alpha < \alpha_{3}] \wedge [ \alpha_{4} > \alpha_{2}] \wedge [\alpha_{5} > \alpha_{3}]\\
\end{cases}
&\quad \quad \text{Case 3}
\\
1 - \bar{F}_{|h_{1}|^{2}}(\zeta_{2}) \bar{F}_{|h_{2}|^{2}}(\zeta_{4}), \quad \quad \begin{cases} 
[\alpha_{2} < \alpha < \alpha_{3}]  \wedge [ \alpha_{5} < \alpha_{2}] \\
[\alpha_{5} < \alpha < \alpha_{3}] \wedge [ \alpha_{5} > \alpha_{2}] \\
\end{cases}
&\quad \quad \text{Case 4}
\\
1, \quad \quad \quad \quad \quad \quad \quad \quad \quad \quad \quad \quad \text{otherwise.}
&\quad \quad \text{Case 5}
\end{cases}
\end{equation}
\noindent\rule{18cm}{0.5pt}
\end{figure*}
Here,
\begin{align}
\bar{F}_{|h_{1}|^{2}}(\zeta_{1}) &= \exp\bigg\{-\frac{\pi_{1}}{(\alpha-\beta(1-\alpha)\pi_{1})\rho_{t}\lambda_{1}}\bigg\}, \\
\bar{F}_{|h_{1}|^{2}}(\zeta_{2}) &= \exp\bigg\{-\frac{\pi_{2}}{(1-\alpha-\alpha\pi_{2})\rho_{t}\lambda_{1}}\bigg\},\\
\bar{F}_{|h_{2}|^{2}}(\zeta_{3}) &= \exp\bigg\{-\frac{\pi_{1}}{(\alpha-(1-\alpha)\pi_{1})\rho_{t}\lambda_{2}}\bigg\}, \\
 \bar{F}_{|h_{2}|^{2}}(\zeta_{4}) &= \exp\bigg\{-\frac{\pi_{2}}{(1-\alpha-\beta\alpha\pi_2)\rho_{t}\lambda_{2}}\bigg\}.
\end{align}

\section{Pair Outage Probability Minimization}
In this section, we obtain optimal power allocation that minimizes pair outage probability. 

\subsection{Problem Formulation}
Observing POP expression as a function of $\alpha$, the POP optimization problem, using \eqref{pair_outage}, can be formulated as
\begin{align}
\mathcal{O}: \underset{\alpha}{\text{minimize}} && P_{o}, &&
\text{s.t.} && (C1): 0<\alpha<1, \nonumber
\end{align}
where $(C1)$ denotes the constraint on PA coefficient. 

\subsection{Solution Methodology}
POP $P_{o}$, given in \eqref{pair_outage}, is a closed-form piece-wise expression. We study each case of $P_{o}$ one by one, and find candidate optimal points to obtain the global-optimal solution. The key result regarding the global-optimal solution of $\mathcal{O}$ is provided through the Lemma $1$.
\begin{Lemma}
The  global-optimal power allocation $\alpha^{*}$ is the feasible optimal point from the set of obtained candidate optimal points that minimizes $P_{o}$, which can be expressed as   
\begin{align}
\alpha^{*}\!\stackrel{\Delta}{=}\!\underset{\alpha \in \{\alpha_{c1}, \alpha_{r1}, \alpha_{r2}, \alpha_{r3}, \alpha_{r4}, \alpha_{c2}\}}{\mathrm{argmin}} P_{o},
\end{align}
where $\alpha_{c1}$, $\alpha_{r1}$, $\alpha_{r2}$, $\alpha_{r3}$, $\alpha_{r4}$ and $\alpha_{c2}$ are the candidate optimal points.
\end{Lemma}

\begin{IEEEproof}
Let us consider each case of the closed-form piecewise expression of $P_{o}$ given in \eqref{pair_outage}.

\textit{Case 1 :} In the first case, when $P_{o}=1 - \bar{F}_{|h_{1}|^{2}}(\zeta_{1}) \bar{F}_{|h_{2}|^{2}}(\zeta_{3})$, the first-order derivative obtained by differentiating $P_{o}$ with respect to $\alpha$ is given as 
\begin{align}\label{case1}
\frac{\text{d}P_o}{\text{d}\alpha}&=-\left(\dfrac{\pi_1s_1}{\lambda_1\rho_{t}\left(\alpha s_1-\beta\pi_1\right)^2}+\dfrac{\pi_1 t_1}{\lambda_2\rho_{t}\left(\alpha t_1-\pi_1\right)^2}\right) \times \nonumber \\ 
&  \mathrm{exp}\bigg\{\!\frac{-\pi_1}{\lambda_1\rho_{t}\left(\alpha s_1-\beta\pi_1\right)}\!-\!\frac{\pi_1}{\lambda_2\rho_{t}\left(\alpha t_1-\pi_1\right)}\!\bigg\}.
\end{align}
wher $s_1=\beta\pi_1+1$ and $t_1=\pi_1+1$.
From \eqref{case1}, we observe that $\frac{\mathrm{d} P_{o}}{\mathrm{d}\alpha}<0$. Therefore, $P_{o}$ is a monotonically decreasing function of $\alpha$ in the given range. Thus, the optimal power allocation is  the corner point of the range of $\alpha$ in case-1. Hence, the optimal solution $\alpha_{c1} = \alpha_{5}$ if $\alpha_{5}<\alpha_{2}$, or $\alpha_{c1} = \alpha_{2}$ if $\alpha_{5}>\alpha_{2}$.

\textit{Case 2 :} In the second case, where $P_{o}=1 - \bar{F}_{|h_{1}|^{2}}(\zeta_{1}) \bar{F}_{|h_{2}|^{2}}(\zeta_{4})$, the derivative $\frac{\text{d}P_o}{\text{d}\alpha}$ is obtained as
\begin{align}\label{case2}
\frac{\text{d}P_o}{\text{d}\alpha} &=\left(\!\dfrac{\pi_2 s_2}{\lambda_2\rho_{t}\left(1-\alpha s_2\right)^2}\!-\!\dfrac{\pi_1\ s_1}{\lambda_1\rho_{t}\left(\alpha s_1-\beta\pi_{1}\right)^2}\!\right) \times \nonumber \\ 
&  \mathrm{exp}\bigg\{\!\frac{-\pi_2}{\lambda_2\rho_{t}\left(1-\alpha s_2\right)}\!-\!\frac{\pi_1}{\lambda_1\rho_{t}\left(\alpha s_1-\beta\pi_1\right)}\!\bigg\}.
\end{align}
where $s_2=\beta\pi_2+1$. We observe that $\frac{\text{d}P_o}{\text{d}\alpha}$ in \eqref{case2} does not indicate any monotonicity. But, $P_{o}$ is monotonically increasing function of $\alpha$ if $\dfrac{\pi_2 s_2}{\lambda_2\rho_{t}\left(1-\alpha s_2\right)^2}>\dfrac{\pi_1\ s_1}{\lambda_1\rho_{t}\left(\alpha s_1-\beta\pi_{1}\right)^2}$, and monotonically decreasing function otherwise. Thus, we can obtain the point of inflection by solving $\dfrac{\pi_2 s_2}{\lambda_2\rho_{t}\left(1-\alpha s_2\right)^2} = \dfrac{\pi_1\ s_1}{\lambda_1\rho_{t}\left(\alpha s_1-\beta\pi_{1}\right)^2}$ which can be simplified as a quadratic equation $m_{1}\alpha^2 + m_{2}\alpha + m_{3} = 0$ where $m_{1} \stackrel{\Delta}{=} r_{1}s_{1} - r_{2}s_{2}, m_{2} \stackrel{\Delta}{=} 2(r_{2} - r_{1}\beta\pi_{1}), m_{3} \stackrel{\Delta}{=}\pi_{1}^2\pi_{2}s_{2}\beta^2\lambda_{1} - \pi_{1}s_{1}\lambda_{2}$ with $r_{1} \stackrel{\Delta}{=} \pi_{2}\lambda_{1}s_{1}s_{2},$ and $r_{2} \stackrel{\Delta}{=} \pi_{1}\lambda_{2}s_{1}s_{2}$. The optimal solutions,  which are the roots of the quadratic equation, are given as
\begin{equation}
 \alpha_{r1},\alpha_{r2}=\frac{- m_{2} \pm \sqrt{m_{2}^2-4m_{1}m_{3}}}{2m_{1}}.
\end{equation}

%\begin{figure*}
%\begin{eqnarray}
%\alpha_{r1}=-\dfrac{\left(B^2P_1P_2-1\right)\sqrt{\left(B^2L_1L_2P_1^2+BL_1L_2P_1\right)P_2^2+\left(BL_1L_2P_1^2+L_1L_2P_1\right)P_2}+\left(B^3L_1P_1^2+B^2L_1P_1\right)P_2^2+\left(\left(B^2L_1-B^2L_2\right)P_1^2+\left(BL_1-BL_2\right)P_1\right)P_2-BL_2P_1^2-L_2P_1}{\left(\left(B^3L_2-B^3L_1\right)P_1^2+\left(B^2L_2-2B^2L_1\right)P_1-BL_1\right)P_2^2+\left(\left(2B^2L_2-B^2L_1\right)P_1^2+\left(2BL_2-2BL_1\right)P_1-L_1\right)P_2+BL_2P_1^2+L_2P_1}, \\
%\alpha_{r2}=\dfrac{\left(B^2P_1P_2-1\right)\sqrt{\left(B^2L_1L_2P_1^2+BL_1L_2P_1\right)P_2^2+\left(BL_1L_2P_1^2+L_1L_2P_1\right)P_2}+\left(-B^3L_1P_1^2-B^2L_1P_1\right)P_2^2+\left(\left(B^2L_2-B^2L_1\right)P_1^2+\left(BL_2-BL_1\right)P_1\right)P_2+BL_2P_1^2+L_2P_1}{\left(\left(B^3L_2-B^3L_1\right)P_1^2+\left(B^2L_2-2B^2L_1\right)P_1-BL_1\right)P_2^2+\left(\left(2B^2L_2-B^2L_1\right)P_1^2+\left(2BL_2-2BL_1\right)P_1-L_1\right)P_2+BL_2P_1^2+L_2P_1}
%\end{eqnarray}
%\noindent\rule{18cm}{0.5pt}
%\end{figure*}

\textit{Case 3 :} In the third case, where $P_{o}=1 - \bar{F}_{|h_{1}|^{2}}(\zeta_{2}) \bar{F}_{|h_{2}|^{2}}(\zeta_{3})$, the derivative of $P_{o}$ with respect to $\alpha$ is given as
\begin{align}\label{case3}
\frac{\text{d}P_o}{\text{d}\alpha}&=\left(\dfrac{\pi_2 t_2}{\lambda_1\rho_{t}\left(1-\alpha t_2\right)^2}-\dfrac{\pi_1 t_1}{\lambda_2\rho_{t}\left(\alpha t_1-\pi_{1}\right)^2}\right) \times \nonumber \\ 
&  \mathrm{exp}\bigg\{\frac{-\pi_2}{\lambda_1\rho_{t}\left(1-\alpha t_2\right)}-\frac{\pi_1}{\lambda_2\rho_{t}\left(\alpha t_1-\pi_1\right)}\bigg\}.
\end{align}
where $t_2=\pi_2+1$. Here also, we observe that $\frac{\text{d}P_o}{\text{d}\alpha}$ given in \eqref{case3} does not show any monotonicity. Therefore, similar to the second case, the point of inflection can be obtained by solving $\dfrac{\pi_2 t_2}{\lambda_1\rho_{t}\left(1-\alpha t_2\right)^2} = \dfrac{\pi_1 t_1}{\lambda_2\rho_{t}\left(\alpha t_1-\pi_{1}\right)^2}$ which is simplified as a quadratic equation $l_{1}\alpha^2 + l_{2}\alpha + l_{3} = 0$ where $l_{1} \stackrel{\Delta}{=} q_{2}t_{1} - q_{1}t_{2}, l_{2} \stackrel{\Delta}{=} 2(q_{1} - q_{2}\pi_{1}), l_{3} \stackrel{\Delta}{=}\pi_{1}^2\pi_{2}t_{2}\lambda_{2} - \pi_{1}t_{1}\lambda_{1}$ with $q_{1} \stackrel{\Delta}{=} \pi_{1}\lambda_{1}t_{1}t_{2},$ and $q_{2} \stackrel{\Delta}{=} \pi_{2}\lambda_{2}t_{1}t_{2}$. In this case, the optimal solutions given as the roots of the quadratic equation, are obtained as
\begin{equation}
 \alpha_{r3},\alpha_{r4}=\frac{- l_{2} \pm \sqrt{l_{2}^2-4l_{1}l_{3}}}{2l_{1}}.
\end{equation}

%\begin{figure*}
%\begin{eqnarray}
%\alpha_{r3}=-\dfrac{\left(P_1P_2-1\right)\sqrt{\left(L_1L_2P_1^2+L_1L_2P_1\right)P_2^2+\left(L_1L_2P_1^2+L_1L_2P_1\right)P_2}+\left(-L_2P_1^2-L_2P_1\right)P_2^2+\left(\left(L_1-L_2\right)P_1^2+\left(L_1-L_2\right)P_1\right)P_2+L_1P_1^2+L_1P_1}{\left(\left(L_2-L_1\right)P_1^2+\left(2L_2-L_1\right)P_1+L_2\right)P_2^2+\left(\left(L_2-2L_1\right)P_1^2+\left(2L_2-2L_1\right)P_1+L_2\right)P_2-L_1P_1^2-L_1P_1},\\
%\alpha_{r4}=\dfrac{\left(P_1P_2-1\right)\sqrt{\left(L_1L_2P_1^2+L_1L_2P_1\right)P_2^2+\left(L_1L_2P_1^2+L_1L_2P_1\right)P_2}+\left(L_2P_1^2+L_2P_1\right)P_2^2+\left(\left(L_2-L_1\right)P_1^2+\left(L_2-L_1\right)P_1\right)P_2-L_1P_1^2-L_1P_1}{\left(\left(L_2-L_1\right)P_1^2+\left(2L_2-L_1\right)P_1+L_2\right)P_2^2+\left(\left(L_2-2L_1\right)P_1^2+\left(2L_2-2L_1\right)P_1+L_2\right)P_2-L_1P_1^2-L_1P_1}
%\end{eqnarray}
%\noindent\rule{18cm}{0.5pt}
%\end{figure*}

\textit{Case 4 :} In the fourth case, where $P_{o}=1-\bar{F}_{|h_{1}|^{2}}(\zeta_{2}) \bar{F}_{|h_{2}|^{2}}(\zeta_{4})$, the derivative $\frac{\text{d}P_o}{\text{d}\alpha}$ is obtained as
\begin{align}\label{case4}
\frac{\text{d}P_o}{\text{d}\alpha}&=\left(\!\dfrac{\pi_2 s_2}{\lambda_2\rho_{t}\left(1-\alpha s_2\right)^2} + \dfrac{\pi_2 t_2 }{\lambda_1\rho_{t}\left(1-\alpha t_2\right)^2}\!\right) \times \nonumber \\ 
& \mathrm{exp}\bigg\{\!\frac{-\pi_2}{\lambda_2\rho_{t}\left(1-\alpha s_2\right)}\!-\!\frac{\pi_2}{\lambda_1\rho_{t}\left(1-\alpha t_2\right)}\!\bigg\}.
\end{align}
which is greater than zero. Thus, $P_{o}$ is a monotonically increasing function of $\alpha$. Therefore, the optimal point, in this case, is given by the lower corner point of considered range. Thus, the optimal solution $\alpha_{c2} = \alpha_{2}$ when $\alpha_{5}<\alpha_{2}$, or $\alpha_{c2} = \alpha_{5}$ when $\alpha_{5}>\alpha_{2}$.

\textit{Case 5 :} In the fifth case, $P_{o}$ is a constant equal to the maximum feasible value, i.e., $P_{o}=1$.  

From the above analysis, we observe that $\alpha_{c1}$ and $\alpha_{c2}$ are two corner points due to monotonically decreasing and increasing nature of $P_{o}$, respectively. $\alpha_{r1}$ and $\alpha_{r2}$ are roots of quadratic equation as explained in Case 2. Similarly, $\alpha_{r3}$ and $\alpha_{r4}$ are the roots of quadratic equation obtained in Case 3. As a result, POP minimization problem has global-optimal solution $\alpha^{*}$, which is the feasible optimal point from set \{$ \alpha_{c1}$, $\alpha_{r1}$, $\alpha_{r2}$, $\alpha_{r3}$, $\alpha_{r4}$, $\alpha_{c2}$\} at which $P_{o}$ is minimum. 
\end{IEEEproof}

\section{Numerical Results}
In this section, we present numerical results where we study the outage performance of the NOMA communication system under various system parameters. Downlink NOMA system is considered with one BS and two users. Near and far user, respectively, are assumed to be located at a distance of $d_{1}=50$ meter and $d_{2}=100$ meter from BS. Noise power is taken as $-90$ dBm with noise signal following Gaussian distribution at both users. Small scale fading is assumed to follow an exponential distribution with a mean value equal to 1 for both the links \cite{basepaper}. We have averaged simulation results over $10^6$ randomly generated channel realizations using Rayleigh distribution at both strong and weak users. The value of $L_{p}$ and $e$, respectively, are set to $1$ and $3$. Besides, we have taken $\beta=0.2$, $\alpha=0.5$, $R_{1}^{\mathrm{th}}=0.1$, $R_{2}^{\mathrm{th}}=0.1$ and $\rho_{t}=60$ dB. 

\subsection{Validation of Analysis}
\begin{figure}[!t]
\centering
\includegraphics[scale=.39]{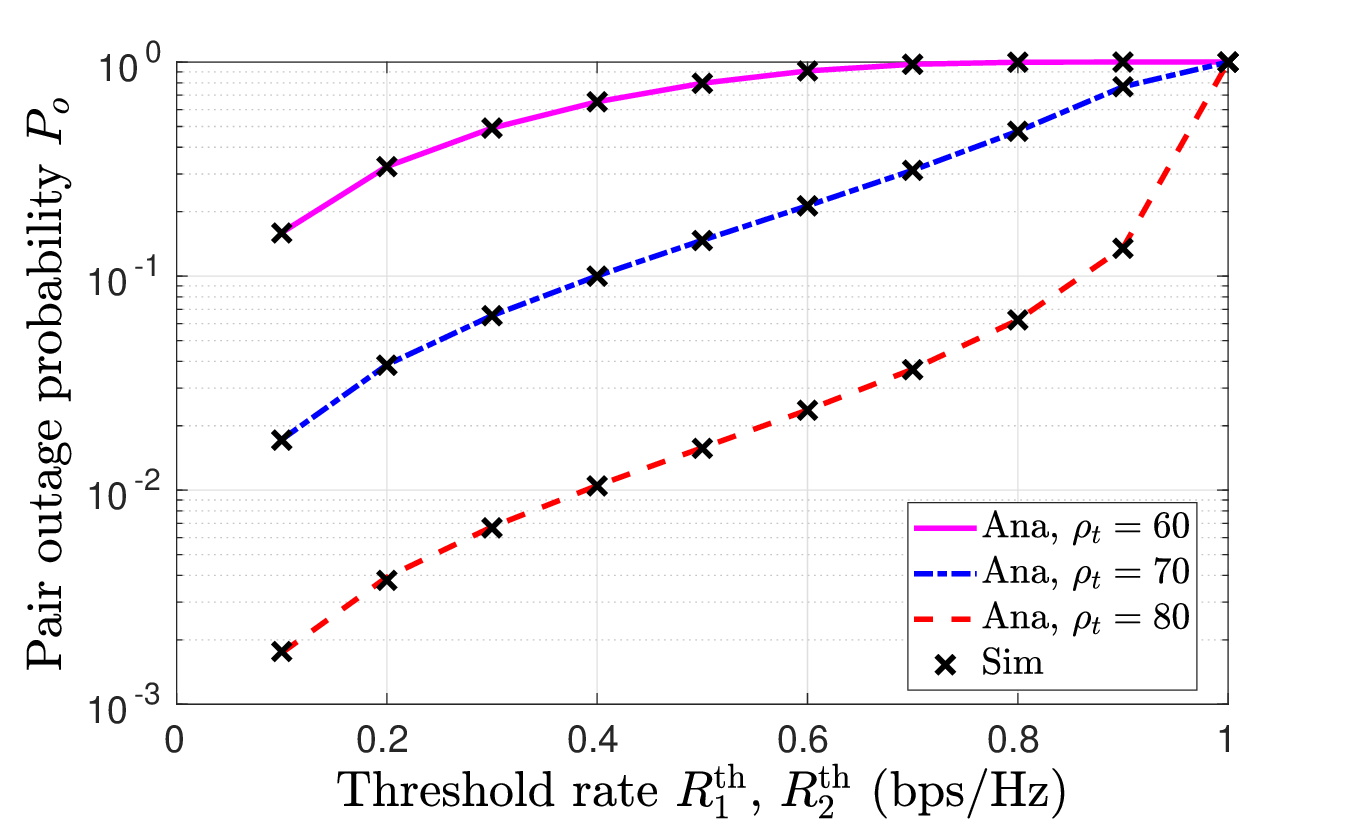}
\vspace{-0.1cm}
\caption{Validation of POP with a variation in threshold data rate requirements at near and far users for various BS transmit SNR values, $\alpha=0.5$.}
\label{fig1}
\end{figure}

Fig. \ref{fig1} presents the validation of the accuracy of the analytical results. The simulation and analytical results are, respectively, marked as `Sim' and `Ana' in the following figures. The validation of POP, i.e., $P_{o}$, is shown with $R_{1}^{\mathrm{th}}$ and $R_{2}^{\mathrm{th}}$ for different values of $\rho_{t}$. The perfect match between simulated and analytical curves confirms the accurateness of POP analysis. Besides, it is observed from the results that with an increase in threshold rates $R_{1}^{\mathrm{th}}$ and $R_{2}^{\mathrm{th}}$, pair outage $P_{o}$  increases. The reason is that an outage happens in the system when the maximum achievable data rate falls below a threshold rate, which clearly means  POP increases on increasing threshold rates at the users. Further, an increase in $\rho_{t}$ results in decrease in POP. This is because, the data rates achieved by users' increase with an increase in the SNR, resulting in a decrease in POP, for given threshold data rates. 

\subsection{Impact of Relative Threshold Data Rate Demands}

\begin{figure}[!t]
\centering
\includegraphics[scale=.39]{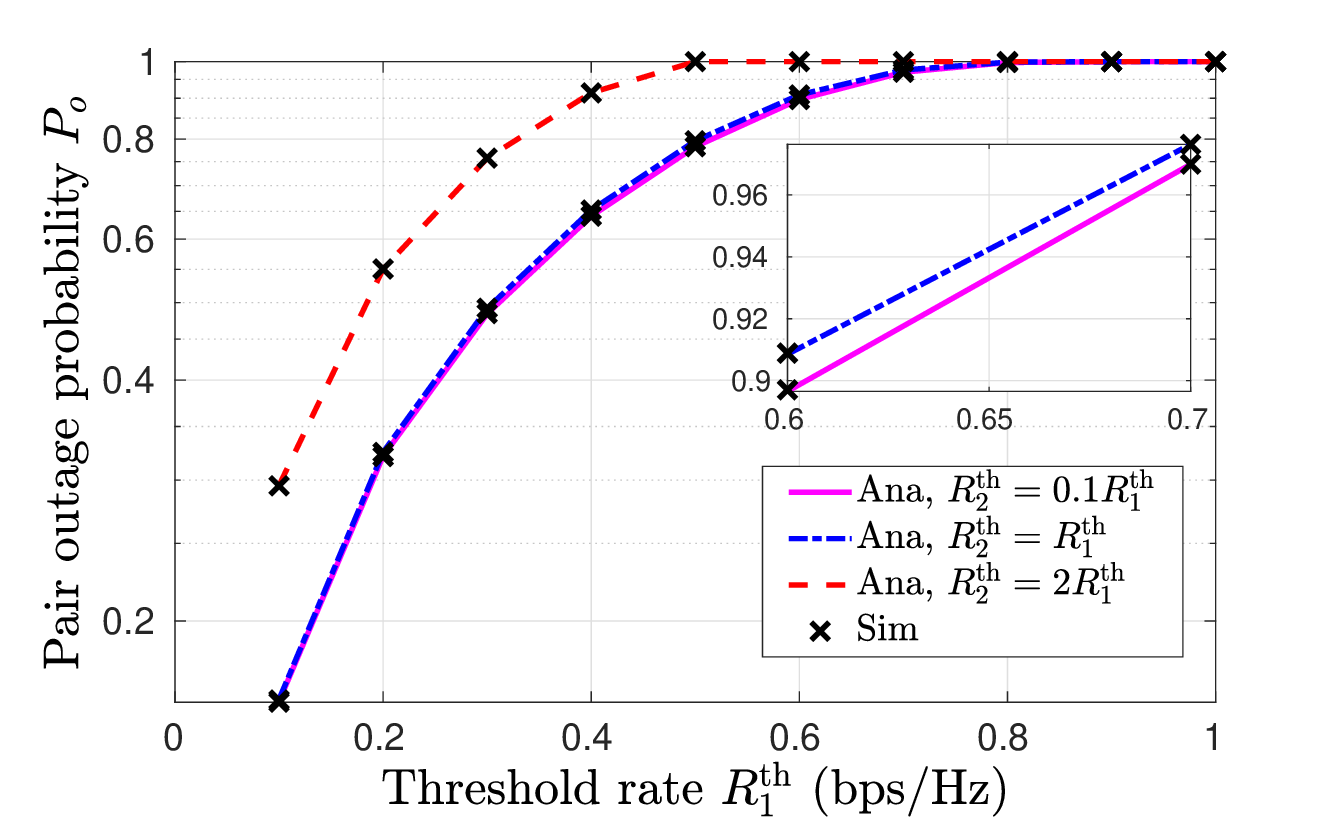}
\vspace{-0.1cm}
\caption{Variation in POP with different threshold data rate requirements at near and far users, $\alpha=0.5$ and $\rho_{t}=60$ dB.}
\label{fig2}
\end{figure}

The impact of relative threshold data rate demands of the near and the far user on the pair outage performance of the system is highlighted through Fig. \ref{fig2}. $P_{o}$ has been plotted versus $R_{1}^{\mathrm{th}}$ for different values of $R_{2}^{\mathrm{th}}$. From the results, it can be observed that the POP performance is better when $R_{1}^{\mathrm{th}} > R_{2}^{\mathrm{th}}$. Note that due to the poorer channel condition of the weak user, the data rate achieved at $U_2$ is always lesser compared  to data rate achieved by $U_1$. In such situation, increasing  $R_{2}^{\mathrm{th}}$ will in turn increase outage probability for $U_2$ which will further increase POP. Therefore, to achieve better pair outage performance,  having the threshold rate pair ($R_{1}^{\mathrm{th}}, R_{2}^{\mathrm{th}}$) with $R_{2}^{\mathrm{th}} < R_{1}^{\mathrm{th}}$ is a good choice.

\subsection{Insights on Optimal Power Allocation}

\begin{figure}[!t]
\centering
\includegraphics[scale=.39]{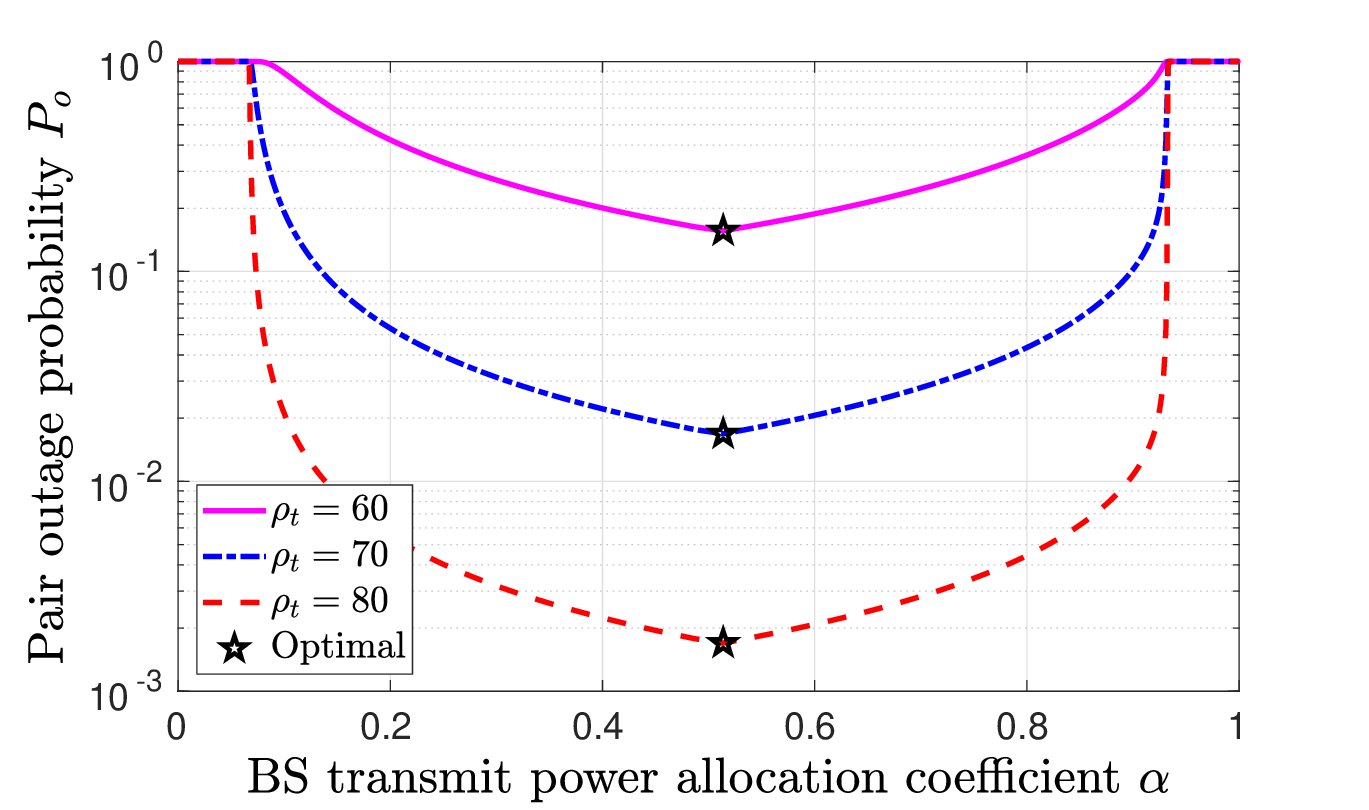}
\vspace{-0.1cm}
\caption{Optimality of POP with power allocation coefficient $\alpha$ associated to near user for different values of BS transmit SNR, $R_{1}^{\mathrm{th}}=R_{2}^{\mathrm{th}}=0.1$.}
\label{fig3}
\end{figure}

\begin{figure}[!t]
\centering
\includegraphics[scale=.39]{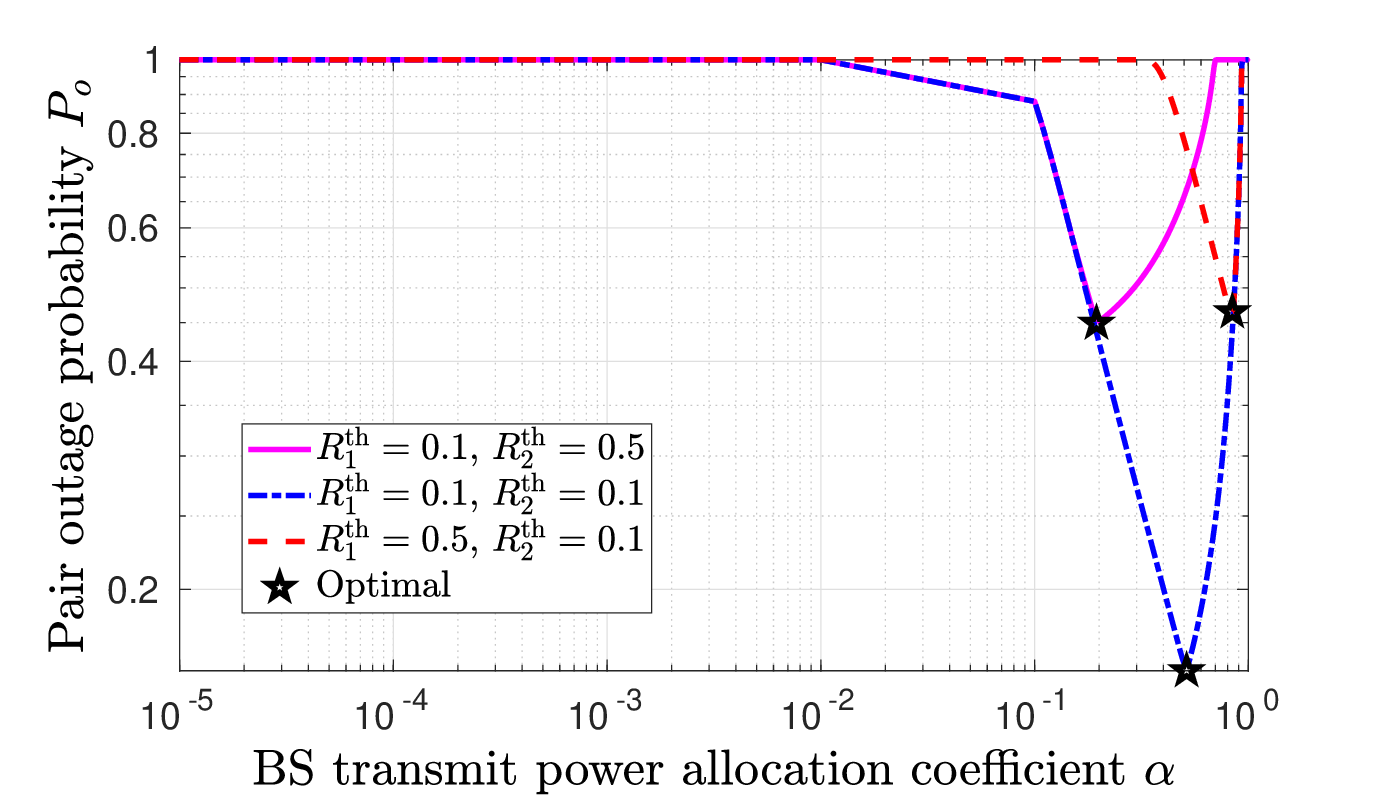}
\vspace{-0.1cm}
\caption{POP versus power allocation coefficient $\alpha$ with different threshold rate requirements at near and far users, $\rho_{t}=60$ dB.}
\label{fig4}
\end{figure}

Both Fig. \ref{fig3} and Fig. \ref{fig4} present the numerical proof of the unique solution of POP  with respect to power allocation $\alpha$.  In Fig. \ref{fig3} and Fig. \ref{fig4}, respectively, we depict $P_{o}$ with $\alpha$ for different values of BS transmit SNR $\rho_{t}$, and threshold rates $R_{1}^{\mathrm{th}}$ and $R_{2}^{\mathrm{th}}$. In both Fig. \ref{fig3} and Fig. \ref{fig4}, the unique optimal solution of power allocation that gives minimum POP can be easily observed. The analytical optimal solution has also been marked to validate the analysis. It is observed from the results that optimal power allocation is independent of the SNR values, which corroborates our analysis presented through Lemma $1$.  However, for the different pair of threshold data rate demands, optimal $\alpha$ varies. Thus, it can be concluded the threshold rate pair plays an important role in the pair outage performance of the system. 

\subsection{Performance Comparison}

\begin{figure}[!t]
\centering
\includegraphics[scale=.4]{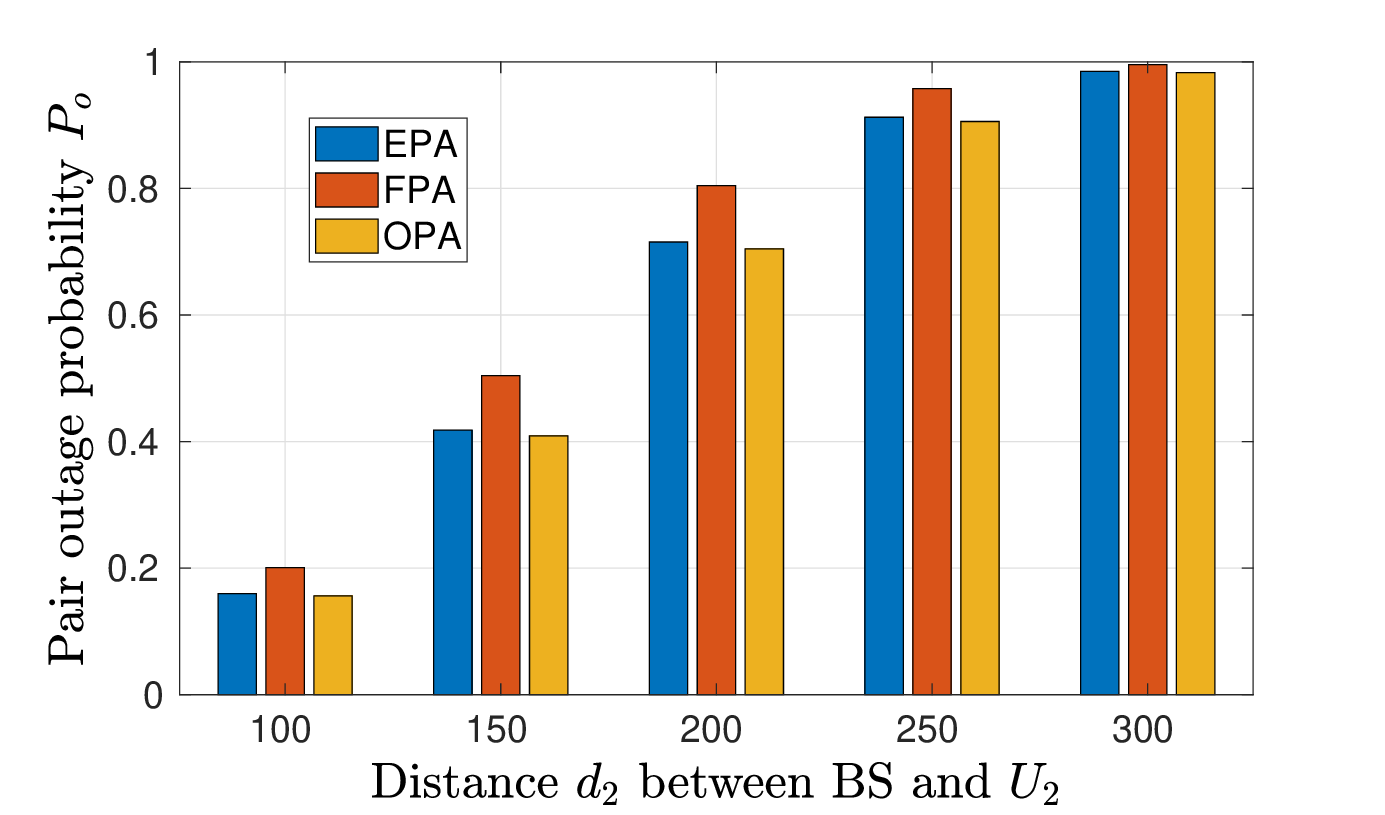}
\vspace{-0.1cm}
\caption{Performance comparison of optimal power allocation with equal power allocation and  fixed power allocation, $d_1 = 50$ meter.}
\label{fig5}
\end{figure}

Finally, Fig. \ref{fig5} demonstrates the performance gain achieved by the optimal solution over two different benchmark schemes. In particular,  we depict the performance comparison of optimal power allocation (OPA) scheme with equal power allocation (EPA) and fixed power allocation (FPA) schemes. In EPA scheme, we have taken $\alpha=0.5$ while in FPA, $\alpha$ is set to 0.4. From the obtained results, it is observed that the optimal power allocation to users  provides considerable improvement in the pair outage performance of the system. The average percentage improvement obtained by OPA over EPA and FPA, are around $1.39 \%$ and $14.60 \%$, respectively. Besides, in Fig. \ref{fig5}, fixing $d_{1}$ as $50$ meter, the effect of variation of $d_{2}$ on POP is also presented. It is observed from the results that with a rise in $d_{2}$, POP increases. This is because on increasing the $d_{2}$, achievable data rate at $U_{2}$ decreases due to which POP of the system increases.

\section{Concluding Remarks}
In this work, we focused on analyzing the pair outage performance of a two-user untrusted NOMA system. In particular, considering imperfect SIC at receivers, we evaluated the closed-form expression of pair outage probability POP for a decoding order that has been proven to be optimal from the perspective of secure communication. We also presented power allocation optimization to minimize the POP. We provided numerical results to validate the accuracy of the derived analytical expressions, and presented the impact of various system parameters on the outage performance. It has been observed that significant performance gains of about $1.39 \%$ and $14.60 \%$, respectively, are achieved by using optimal power allocation compared to equal and fixed power allocations. In future, we wish to focus on generalizing the POP study for all possible decoding orders in a multi-user untrusted NOMA scenario.
\bibliographystyle{IEEEtran}
\bibliography{ref}
\end{document}